\documentclass[twocolumn,prl,aps,amsfonts,amssymb,amsmath,showpacs]{revtex4}
\usepackage{graphicx}
\usepackage{dcolumn}

\graphicspath{{figures/}}
\setkeys{Gin}{width=\linewidth}

\begin{document}
\title{Imaging Electron Wave Functions Inside Open Quantum Rings}

\author{F. Martins$^{1}$, B. Hackens$^{1,2}$, M. G. Pala$^{3}$, T. Ouisse$^{1}$, H. Sellier$^{1}$, \\X. Wallart$^{4}$, S. Bollaert$^{4}$, A. Cappy$^{4}$, J. Chevrier$^{1}$, V. Bayot$^{2}$ and  S. Huant$^{1}$}
\affiliation{$^{1}$ Institut N\'eel, CNRS and Universit\'e Joseph Fourier, BP 166, 38042 Grenoble cedex 9, France\\
$^{2}$ CERMIN, DICE Lab, UCL, B-1348 Louvain-la-Neuve, Belgium\\
$^{3}$ IMEP-MINATEC (UMR CNRS/INPG/UJF 5130), Grenoble, France\\
$^{4}$ IEMN, Cit\'e scientifique, Villeneuve dÕAscq, France\\ 
}

\date{\today}

\begin{abstract}
Combining Scanning Gate Microscopy (SGM) experiments and simulations, we demonstrate low temperature imaging of electron probability density $|\Psi|^{2}(x,y)$  in embedded mesoscopic quantum rings (QRs). The tip-induced conductance modulations share the same temperature dependence as the Aharonov-Bohm effect, indicating that they originate from electron wavefunction interferences. Simulations of both $|\Psi|^{2}(x,y)$ and SGM conductance maps reproduce the main experimental observations and link fringes in SGM images to  $|\Psi|^{2}(x,y)$. 
\end{abstract}

\pacs{73.21.La,73.23.Ad,03.65.Yz,85.35.Ds}
\maketitle

Thanks to the scanning tunnelling microscope (STM), remarkable precision has been achieved in the local scale imaging of surface electron systems. Only a few years after the STM invention, electron interferences could be visualized in real space inside artificially confined surface structures, the "quantum corrals"~\cite{Crommie_sc93}. However, since they rely on the measurement of a current between a tip and the sample, STM techniques are useless when the system of interest is buried under an insulating layer, as in two-dimensional electron gases (2DEGs) confined in semiconductor heterostructures. To circumvent the obstacle, a new method was developed: the Scanning Gate Microscopy (SGM). SGM consists in mapping the conductance of the system as the polarized tip, acting as a flying nano-gate, scans at a constant distance above the 2DEG. SGM gave many valuable insights into the physics of quantum point contacts (QPCs)~\cite{Topinka_science00}, Coulomb-blockaded quantum dots~\cite{pioda:216801}, magnetic focusing~\cite{westervelt_focusing}, carbon nanotubes~\cite{bachtold:6082}, open billiards~\cite{crook_prl03} and 2DEGs in the quantum Hall regime \cite{Finkelstein_2000}. 

In some cases, the mechanism of SGM image formation is readily understandable. For example, in the vicinity of a QPC \cite{Topinka_science00}, coherent electron flow is imaged due to multiple reflections and interferences of electrons bouncing between the QPC and the tip-induced depleted region. 
In comparison, the situation seems more complex when the tip scans directly over an open mesoscopic billiard~\cite{crook_prl03}: the tip perturbation extends over the whole system of interest, so that \emph{all} semi-classical trajectories are modified. The mechanisms that link conductance maps to the properties of unperturbed electrons still need to be clarified.  
Recently, we showed that SGM images in the vicinity of a QR allow direct observation of iso-phase lines for electrons in an electrostatic Aharonov-Bohm (AB) experiment~\cite{hackens_Nphys06}.

In this Letter, we discuss SGM images obtained as the tip scans directly over coherent quantum rings (QRs). Experimentally, we find that the amplitude of conductance modulations shares a common temperature dependence with the Aharonov-Bohm effect, a direct evidence that SGM probes the quantum nature of electrons. On the other hand, we perform quantum mechanical simulations of SGM experiments. First, the amplitude of conductance fringes is found to evolve linearly at low perturbation amplitude, both in experiments and simulations. Second, we observe a direct correspondence between simulated SGM data and simulations of  the electron probability density $|\Psi|^{2}(x,y,E_{\rm F})$. We deduce that, in this linear regime, SGM reliably maps $|\Psi|^{2}(x,y,E_{\rm F})$ in coherent QRs.
\begin{figure}
\begin{center}
\includegraphics[width=\columnwidth]{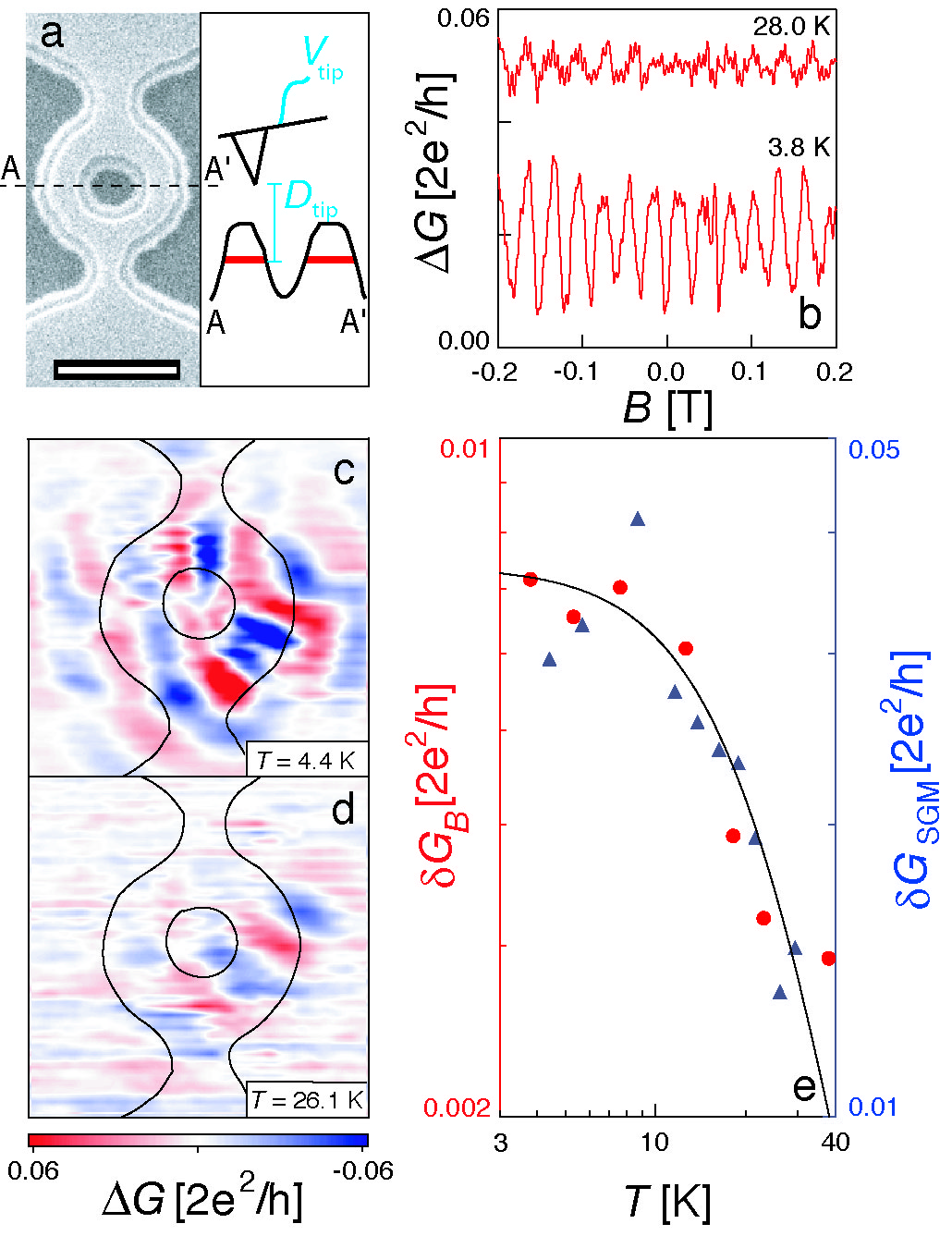}
\caption{(Color online) (a) Left : electron micrograph of sample R2. The white bar represents 500 nm. Right : Scheme of the SGM experiment (side view). The black line represents the profile of the quantum ring along A-A$^\prime$, the red line represents the 2DEG. (b)  $\Delta G$ vs $B$ at 3.8 K and 28.0 K
 in R2. (c-d)  $\Delta G(x,y)$ maps measured on sample R2 at $V_{\mathrm{tip}}= 0 \mathrm V$, 
 $D_{\mathrm{tip}} = 50$ nm, $B = 0$ T and $T = 4.4$ and $26.1$ K, respectively. 
 (e) Red circles : $\delta G_{B}$ vs $T$ (left axis); blue triangles : 
 $\delta G_{\mathrm{SGM}}$ vs $T$ (right axis). The black line is a guide to the eye.}
\label{fig:fig1}
\end{center}
\end{figure}

We fabricated two QRs, samples R1 and R2, from an InGaAs/InAlAs heterostructure using electron beam lithography and wet etching~\cite{sampleR1}.  The 2DEG is located 25 nm below the surface, and its low temperature ($T$) electron density and mobility are $2 \times 10^{16}~\mathrm{m^{-2}}$ and $10~\mathrm{m^{2}/Vs}$, respectively.
Fig.~\ref{fig:fig1}(a) shows an electron micrograph of R2 and a scheme of our experimental setup. The ring inner and outer (lithographic) diameters are 240 and 580 nm, respectively (210 and 600 nm in R1).
At $T=4.2$ K and after illumination, the conductances $G$ of R1 and R2 are 5.9 and 2.6$\times2~\mathrm{e}^{2}/\mathrm{h}$, respectively, \emph{i.e.} several quantum modes are populated in the device openings. 

Figure~\ref{fig:fig1}(b) plots the low-$T$ conductance $\Delta G$ of R2 as a function of the magnetic field $B$ after subtraction of a slowly varying background. Periodic oscillations of  $\Delta G$ vs $B$ are clearly visible at 3.8 K and 28.0 K. They are the signature of phase shifts between electron wavefunctions transmitted through both arms of the QR, \emph{i.e.} the Aharonov-Bohm effect. The damping of AB oscillations observed as $T$ increases [Fig.~\ref{fig:fig1}(b)] is related to the decay of the electron coherence time $\tau_{\phi}$. Nevertheless, the persistence of AB oscillations  guarantees that transport is at least partially coherent up to $T\sim$ 30~K.

We then perform scanning gate microscopy in this coherent regime of transport. Fig.~\ref{fig:fig1}(a) illustrates our imaging technique~: a voltage $V_{\mathrm{tip}}$ is applied on the tip, which scans along a plane parallel to the 2DEG at a typical tip-2DEG distance $D_{\mathrm{tip}}$ = 50 nm. Due to the capacitive tip-2DEG coupling, a local perturbation is generated in the potential experienced by electrons within the QR, which, in turn, alters their transmission through the device. By recording the QR conductance as the tip is scanned over the QR and its vicinity, we build a conductance map $G(x,y)$ that reflects changes in electron transmission.

\begin{figure}
\begin{center}
\includegraphics[width=\columnwidth]{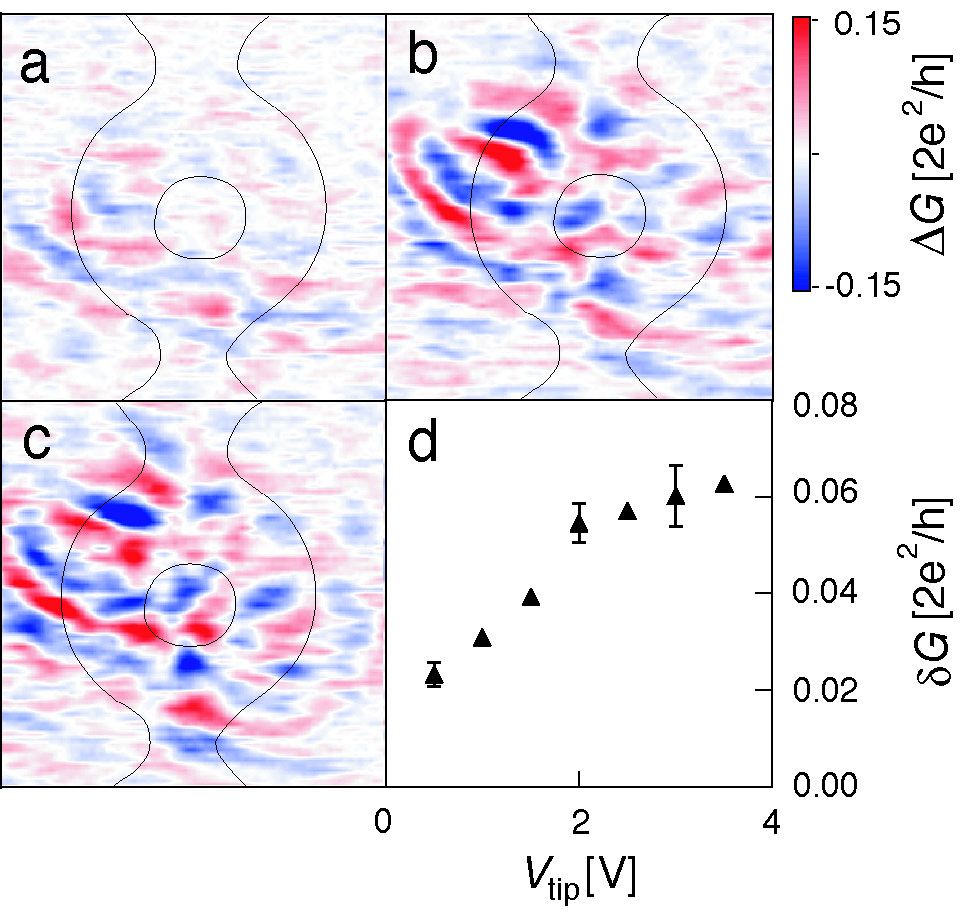}
\caption{(Color online) (a-c) $\Delta G(x,y)$ measured on R1 at $D_{\rm tip} = 50$ nm, $T = 4.2$ K, $B = 2$T and $V_{\rm tip} = 0.5$, 2.5 and 3.5 V, respectively. (d) $\delta G_{\rm SGM}$ vs $V_{\rm tip}$ in R1.}
\label{fig:fig2}
\end{center}
\end{figure}

The measured $G$ maps reveal a rich pattern of conductance fringes superimposed on a broad and slowly varying background, insensitive to temperature and to $B$~\cite{hackens_Nphys06}, indicating that it is not related to quantum transport. We remove the background contribution by high-pass filtering the raw $G$ maps~\cite{hackens_Nphys06}.  Figs.~\ref{fig:fig1}(c-d) show such filtered $\Delta G$ maps measured on R2 at $T=4.4$ and $26.1~\mathrm{K}$. These SGM images are not symmetric, most probably due to unavoidable asymmetry in the QR geometry. Moreover, $\Delta G$ fringes are not limited to the area of the QR: even when the tip scans away from the QR, it perturbs the QR potential and affects its conductance. In an earlier work \cite{hackens_Nphys06}, we showed that concentric $\Delta G$ fringes observed outside the QR area originate from the electrostatic AB effect, and correspond to iso-phase lines for electrons. Here, we focus on the $\Delta G$ fringes measured when the tip scans directly over the QR. 

In Fig.~\ref{fig:fig1}(c-d), we note a decay of $\Delta G$ fringe amplitude with increasing $T$, but the fringes pattern remains essentially unchanged. If we define $\delta G_{\mathrm{SGM}}$, the standard deviations of the filtered conductance maps, calculated on a 400-nm diameter circle centered on the QR, and $\delta G_{B}$, the standard deviation of AB oscillations in $G$ vs $B$, we can compare the $T$-dependences of SGM images and AB effect [Fig.~\ref{fig:fig1}(e)]. $\delta G_{B}$ and $\delta G_{\mathrm{SGM}}$ clearly follow the same $T$-dependence: a strong decay above $T\sim 10$ K, and a saturation at lower $T$, consistent with the intrinsic saturation of $\tau_{\phi}$ previously reported in similar confined systems~\cite{hackens_PRL05}. This suggests that the AB effect and the central pattern of fringes in $\Delta G$ maps are intimately related and find a common origin in electron wavefunctions interferences. The question is then: do $\Delta G$ maps bear spatial informations on electron wavefunctions~?

To investigate further on the SGM imaging mechanism, we need to examine the influence of a key ingredient: the perturbing potential~\cite{gildemeister-2007}. Experimentally, a way to control this parameter is to change the tip voltage. Fig.~\ref{fig:fig2}(a-c) show a sequence of $\Delta G$ maps measured at $V_{\mathrm{tip}}=0.5$, 2.5 and 3.5 V on R1. Clearly, the amplitude of conductance fringes increases with $V_{\mathrm{tip}}$. Furthermore, we note a strong similarity between the central patterns in successive $\Delta G$ maps. This contrasts with aperiodic conductance fluctuations observed in most mesoscopic systems when changing gate voltages \cite{Krafft:2001lr, crook_prl03}, as well as with the concentric fringes observed on the same QR sample, but away from its central region~\cite{hackens_Nphys06}. Fig.~\ref{fig:fig2}(d), showing the fringe amplitude $\delta G_{\rm SGM}$ vs $V_{\mathrm{tip}}$, confirms the absence of fluctuations and the smooth evolution of fringes amplitude: $\delta G_{\rm SGM}$ increases linearly up to $V_{\mathrm{tip}}=2~\mathrm{V}$, with a somewhat weaker dependence at larger $V_{\mathrm{tip}}$.


\begin{figure}
\begin{center}
\includegraphics[width=\columnwidth]{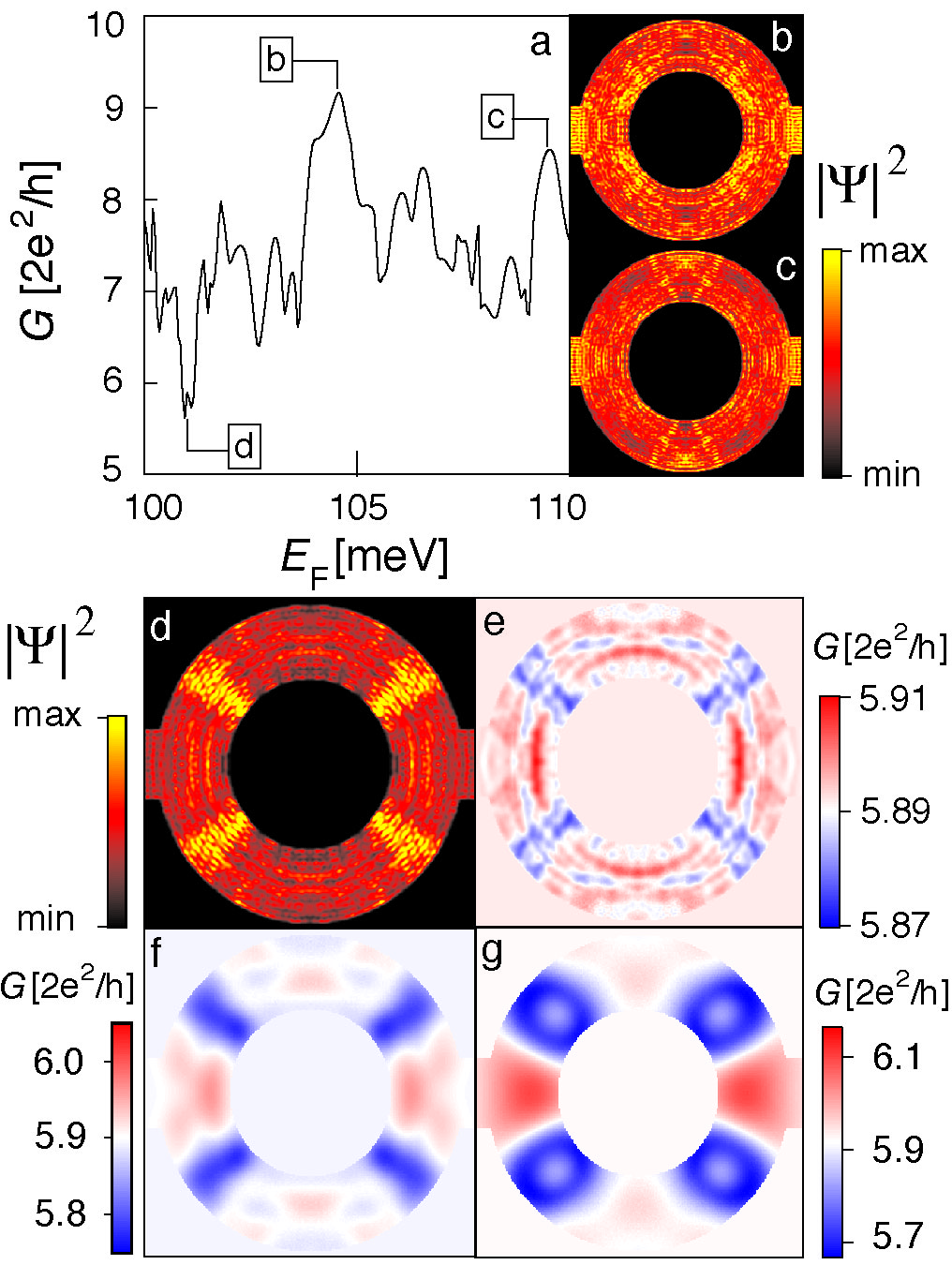}
\caption{(Color online) (a) Calculated $G$ vs $E_{\rm F}$ in a QR with inner and 
outer radii of 140 and 265 nm. (b-d) Electron probability density $|\Psi|^2(x,y,E_{\rm F})$ 
in a QR at $E_{\rm F}=104.6$, 109.5 and 101 meV, respectively. (e-g) Simulated $G(x,y)$ 
for $E_{\rm F}=101$\ meV, $V_{\rm m}=E_{\rm F}/200$ and $\sigma =$ 5, 20 and 40 nm, respectively.}
\label{fig:fig3}
\end{center}
\end{figure}

This striking behavior is now addressed by quantum mechanical simulations where we compute device conductance and local density of states (LDOS) at the Fermi energy $|\Psi|^2(x,y,E_{\rm F})$ in the scattering matrix formalism and using the Landauer-B\"{u}ttiker formula, with the same method as in Ref.~\cite{pala_PRB04}. The geometry of the simulated QR is that of sample R1, taking the depletion length at device edges into account ($\sim$ 35 nm), $\it{i.e.}$ the inner and outer radii of the ring are 140 nm and 265 nm, respectively, and the width of the ring's openings is 120 nm. The ring is subdivided into slices perpendicular to $x$-axis (propagating direction) between both openings in the QR. The Schr\"{o}dinger equation is numerically solved in each slice and, by imposing the continuity of the wavefunction and of the current density at the interface, the scattering matrix between two slices is obtained as prescribed in Ref.~\cite{macucci_PRB92}. The overall scattering matrix is computed by composing the matrices of all slices. 

When $E_{\rm F}$ is varied, both the configuration of resonant states within the QR and their coupling to the reservoirs change \cite{akis_PRL97}. As a consequence, the calculated conductance $G$ exhibits fluctuations as a function of $E_{\rm{F}}$, one of the hallmarks of transport in open mesoscopic systems. This is illustrated in Fig.~\ref{fig:fig3}(a), showing the calculated $G$ as a function of $E_{\rm F}$ around the measured $E_{\rm F}^{\rm 2DEG}$ value in the unpatterned heterostructure. Figs.~\ref{fig:fig3}(b-d) show typical patterns of $|\Psi|^2(x,y,E_{\rm F})$ in our QR, for $E_{\rm F}= 104.6$, 109.5 and 101.0 meV, respectively. Small-scale concentric oscillations are visible within the whole QR area in Figs.~\ref{fig:fig3}(b-d), whose characteristic spatial periodicity is related to the Fermi wavelength $\lambda_{\rm F}$. On a scale larger than $\lambda_{\rm F}$, $|\Psi|^2(x,y,E_{\rm F})$ is rather homogeneous in Figs.~\ref{fig:fig3}(b-c), while it exhibits four strong radial fringes in Figs.~\ref{fig:fig3}(d).

In analogy with the experiment, we included in our simulation a moving perturbation potential mimicking the tip effect, and then calculated the conductance of the QR for each position of the perturbation. Fig.~\ref{fig:fig3}(e) shows such a simulated conductance map $G(x,y)$, obtained for $E_{\rm F}=101$ meV, using the Gaussian potential $V(x,y)=V_{\rm m}*\mathrm{exp}\left\{-\left[\frac{(x-x_{0})^{2}+(y-y_{0})^{2}}{2*\sigma^{2}}\right]\right\}$ as the moving perturbation \cite{shape}, with $V_{\rm m} = E_{\rm F}/200$, $\sigma = 5$ nm and $(x_{0},y_{0})$ the local position of the tip. Most importantly, a careful examination of Fig.~\ref{fig:fig3}(d-e) reveals that all the features visible in the simulated $|\Psi|^2(x,y,E_{\rm F})$ are also visible in the calculated $G$ map. This striking correspondence reveals that SGM can in principle be used to map the unperturbed electron probability density~\cite{maxi}. 

\begin{figure}
\begin{center}
\includegraphics[width=\columnwidth]{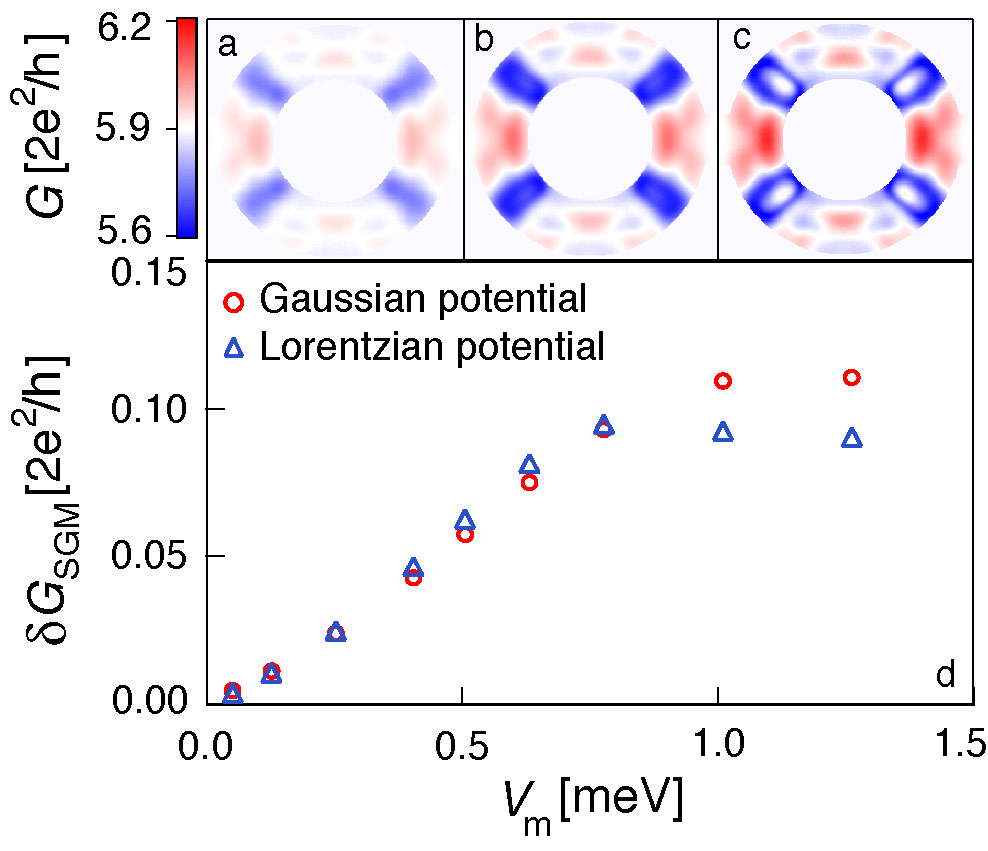}
\caption{(Color online) (a-c) Simulated $G$ maps, calculated 
for $E_{\rm F}$= 101 meV, $\sigma = 20$ nm and $V_{\rm m} =0.5$, 
1.0 and 1.5 meV, respectively. (d) Fringe amplitude $\delta G_{\rm SGM}$ 
on simulated $G$ maps as a function of $V_{\rm m}$, with $\sigma = 20$ nm.}
\label{fig:fig4}
\end{center}
\end{figure}

Enlarging now the width of the perturbing potential causes the smallest SGM features to disappear [Figs.~\ref{fig:fig3}(f-g)]. As $\sigma$ overcomes $\lambda_{\rm F}$, concentric fringes completely vanish. Nevertheless, at the scale of radial fringes, the correspondence between $G(x,y)$ and $|\Psi|^2(x,y,E_{\rm F})$ is maintained [Fig.~\ref{fig:fig3}(f)] until $\sigma \gtrsim 40$ nm where the aspect of the four radial fringes changes (Fig.~\ref{fig:fig3}(g)). Most importantly, the size of the smallest features in the simulated $G$ maps is roughly correlated to $\sigma$. Based on Figs.~\ref{fig:fig2}(a-c), this allows us to infer a lower bound for $\sigma \sim$ 25 nm.

To come closer to a complete description of our experiments, we now examine the effect of the perturbation amplitude. As we increase $V_{\rm m}$, keeping $\sigma= 20$ nm, we observe that the SGM fingerprint remains qualitatively independent of $V_{\rm m}$, and that its amplitude increases linearly with $V_{\rm m}$ [Figs.~\ref{fig:fig4}(a-c)]; a behaviour consistent with the observations related in Figs.~\ref{fig:fig2}(a-c). Above $V_{\rm m}\sim 0.8$ meV, qualitative changes in the SGM pattern appear together  with a deviation from the linear evolution of $\delta G_{\rm SGM}$ vs $V_{\rm m}$ [Fig.~\ref{fig:fig4}(c)], a value for which differences between $G(x,y)$ and $|\Psi|^2(x,y,E_{\rm F})$ start to emerge. Fig.~\ref{fig:fig4}(d) also shows that the evolution of $\delta G_{\rm SGM}$ vs $V_{\rm m}$ is weakly affected by the exact shape of the perturbation, \emph{e.g.} Lorentzian or Gaussian. The consistency between the simulated behavior of $\delta G_{\rm SGM}$ vs $V_{\rm m}$ and the experimental data in Fig.~\ref{fig:fig2}(d) leads us to conclude that, at low $V_{\rm tip}$ (below $\sim$ 2V), the central part of our SGM maps directly reveals the main structure of $|\Psi|^2(x,y,E_{\rm F})$ in our quantum rings. This means that we can attribute the pattern of conductance fringes in Fig.~\ref{fig:fig2}(a-c) and Fig.~\ref{fig:fig1}(c-d) to $|\Psi|^2(x,y,E_{\rm F})$ of electrons in the quantum ring. Furthermore, simulations of $|\Psi|^2(x,y,E_{\rm F})$ in asymmetric quantum rings yield asymmetric structures \cite{pala_future}, which gives us an explanation for the asymmetry observed in the experimental $\Delta G$ maps.

In summary, we observed a pattern of fringes in conductance maps obtained by scanning gate microscopy on quantum rings. Using quantum mechanical simulations of the electron probability density, including the perturbing potential of the tip, we could reproduce the main experimental features and demonstrate the relationship between conductance maps and electron probability density maps. Hence, one can view SGM as the analog of STM for imaging the electronic LDOS in open mesoscopic systems buried under an insulating layer, or the counterpart of the near-field scanning optical microscope that images the photonic LDOS in confined nanostructures~\cite{colas}. 


F. M. is funded by FCT (Portugal) and B. H. by the EU (Marie Curie IEF) and FNRS. This work has been supported by the Communaut\'e Fran\c{c}aise de Belgique (Actions de Recherche Concert\'ees), by FRFC grant no. 2.4502.05, by the Belgian Science Policy 
(Interuniversity Attraction Pole Program PAI), by the Action Concert\'ee Nanoscience (French Ministry for Education and Research) and by the Institut de Physique de la Mati\`ere Condens\'ee, Grenoble.
F. M. and B. H. contributed equally to this work.

\bibliographystyle{prsty}

\begin{thebibliography}{10}

\bibitem{Crommie_sc93}
M.~F. Crommie, C.~P. Lutz, and D.~M. Eigler, Science {\bf 262},  218  (1993).

\bibitem{Topinka_science00}
M.~A. Topinka,  B.~J. LeRoy,  S.~E. J.~Shaw,  E.~J. Heller,  R.~M. Westervelt,  K.~D. Maranowski,  and A.~C. Gossard,  Science {\bf 289},  2323  (2000); 
M.~A. Topinka, B.~J. LeRoy, R.~M. Westervelt, S.~E.~J. Shaw, R. Fleischmann, E.~J. Heller, K.~D. Maranowski, and A.~C. Gossard, Nature {\bf 410},  183  (2001); 
N. Aoki, C.~R. Da~Cunha, R. Akis, D.~K. Ferry, and Y. Ochiai, Appl. Phys. Lett. {\bf 87},  223501  (2005).

\bibitem{pioda:216801}
A. Pioda, S. Kicin, T. Ihn, M. Sigrist, A. Fuhrer, K. Ensslin, A. Weichselbaum, S.~E. Ulloa, M. Reinwald and W. Wegscheider, Phys. Rev. Lett. {\bf 93},  216801  (2004);
P. Fallahi, A.~C. Bleszynski,  R.~M. Westervelt,  J.  Huang,  J.~D. Walls, E.~J. Heller,  M. Hanson, and A.~C. Gossard, Nano Letters {\bf 5},  223  (2005).

\bibitem{westervelt_focusing}
R. Crook, C.~G. Smith, M.~Y. Simmons, and D.~A. Ritchie, Phys. Rev. B {\bf 62}, 005174 (2000);
K.~E. Aidala, R.~E. Parrott, T. Kramer, E.~J. Heller, R.~M. Westervelt, M.~P. Hanson and A.~C. Gossard, Nature Phys. {\bf 3},  464  (2007).

\bibitem{bachtold:6082}
A. Bachtold, M.~S. Fuhrer, S. Plyasunov, M. Forero, E.~H. Anderson, A. Zettl, and P.~L. McEuen, Phys. Rev. Lett. {\bf 84},  6082  (2000).

\bibitem{crook_prl03}
R. Crook, C.~G. Smith, A.~C. Graham, I. Farrer, H.~E. Beere, and D.~A. Ritchie, Phys. Rev. Lett. {\bf 91},  246803  (2003).

\bibitem{Finkelstein_2000}
G. Finkelstein, P.~I. Glicofridis, R.~C. Ashoori, and M. Shayegan, Science {\bf
  289},  90  (2000).

\bibitem{hackens_Nphys06}
B. Hackens, F. Martins, T. Ouisse, H. Sellier, S. Bollaert, X. Wallart, A. Cappy, J. Chevrier, V. Bayot, and S. Huant, Nature Phys. {\bf 2},  826  (2006).

\bibitem{sampleR1}
Data from a first set of measurements on R1 were presented in Ref.~\cite{hackens_Nphys06}. Here, we present new data on R1, focusing on the central part of the QR.

\bibitem{hackens_PRL05}
B. Hackens, S. Faniel, C. Gustin, X. Wallart, S. Bollaert, A. Cappy, and V. Bayot, Phys. Rev. Lett. {\bf 94},  146802  (2005).


\bibitem{gildemeister-2007}
A.~E. Gildemeister {\it et~al.}, Phys. Rev. B {\bf 75}, 195338 (2007).

\bibitem{Krafft:2001lr}
B. Krafft, A. Forster, A. van~der Hart, and T. Schapers, Physica E (Amsterdam)
  {\bf 9},  635  (2001).

\bibitem{pala_PRB04}
M.~G. Pala and G. Iannaccone, Phys. Rev. B {\bf 69},  235304  (2004).

\bibitem{macucci_PRB92}
M. Macucci and K. Hess, Phys. Rev. B {\bf 46},  15357  (1992).


\bibitem{akis_PRL97}
R. Akis, D.~K. Ferry, and J.~P. Bird, Phys. Rev. Lett. {\bf 79},  123  (1997).

\bibitem{shape}
In SGM measurements on isolated quantum dots~\cite{gildemeister-2007}, 
the shape of the tip potential was found to be Lorentzian. In our simulated $G$ maps, we did not observe significant differences between Gaussian and Lorentzian perturbing potential of comparable amplitude and lateral extension.

\bibitem{maxi}
In the case of a repulsive potential and for $E_{\rm F}=101$ meV (as in our simulations), a minimum is observed in $G$ maps for a maximum in $|\Psi|^2$. For an attractive potential, a maximum in $|\Psi|^2$ corresponds to a maximum in $G$ maps.

\bibitem{pala_future}
M.~G. Pala {\it et~al.}, in preparation.

\bibitem{colas}
G. Colas des Francs, C. Girard, J.-C. Weeber, C. Chicane, T. David, A. Dereux, and D. Peyrade, Phys. Rev. Lett. {\bf 86},  4950  (2001).

\end{thebibliography}



\end{document}